\documentclass[aps,prd,twocolumn,floatfix,nofootinbib,superscriptaddress,tightenlines]{revtex4}
\usepackage{amsmath}
\usepackage{lipsum}
\usepackage{amssymb}
\usepackage{amsthm}
\usepackage{bbold}
\usepackage{dcolumn}
\usepackage{epsfig}
\usepackage{graphics}
\usepackage{graphicx}
\usepackage{longtable}
\usepackage{color}
\usepackage{epstopdf}
\usepackage{xspace}
\usepackage{cancel}
\usepackage{nicefrac}
\usepackage[colorlinks=true, pdfstartview=FitV, linkcolor=purple, citecolor= purple,urlcolor=blue]{hyperref}
\usepackage{lipsum}
\usepackage{slashed}

\definecolor{darkgreen}{rgb}{0,0.5,0}
\definecolor{purple}{rgb}{0.5,0,0.5}
\definecolor{nblue}{rgb}{0.0,0.0,0.50}
\definecolor{scarlet}{rgb}{1.0,0.2,0}

\newcommand{\dd}{\ensuremath{\mathrm{d}}}
\em
\begin{document}

\title{Electron-Photon Vertex and Dynamical Chiral Symmetry Breaking in Reduced QED: An Advanced Study of Gauge Invariance}

\author{L. Albino}
\email{albino.fernandez@umich.mx}
\affiliation{Instituto de F\'isica y Matem\'aticas, Universidad Michoacana de San Nicol\'as de Hidalgo, Morelia, Michoac\'an 58040, M\'exico.}

\author{A. Bashir}
\email{adnan.bashir@umich.mx}
\affiliation{Instituto de F\'isica y Matem\'aticas, Universidad Michoacana de San Nicol\'as de Hidalgo, Morelia, Michoac\'an 58040, M\'exico.}

\author{A.J. Mizher}
\email{ana.mizher@unesp.br}
\affiliation{Instituto de F\'isica Te\'orica, Universidade Estadual Paulista, Rua Dr. Bento Teobaldo Ferraz, 271-Bloco II, 01140-070, S\~{a}o Paulo, SP, Brazil.}
\affiliation{Centro de Ciencias Exactas, Universidad del B\'io-B\'io. Avda. Andr\'es Bello 720, Casilla 447, 3800708, Chill\'an, Chile.}

\author{A. Raya}
\email{alfredo.raya@umich.mx}
\affiliation{Instituto de F\'isica y Matem\'aticas, Universidad Michoacana de San Nicol\'as de Hidalgo, Morelia, Michoac\'an 58040, M\'exico.}
\affiliation{Centro de Ciencias Exactas, Universidad del B\'io-B\'io. Avda. Andr\'es Bello 720, Casilla 447, 3800708, Chill\'an, Chile.}

\date{\today}

\begin{abstract}

We study the effect of a refined electron-photon vertex on the dynamical breaking of chiral symmetry in reduced quantum electrodynamics. We construct an educated {\em ansatz} for this vertex which satisfies the required discrete symmetries under parity, time reversal and charge conjugation operations. Furthermore, it reproduces its asymptotic perturbative limit in the weak coupling regime and ensures the massless electron propagator is multiplicatively renormalizable in its leading logarithmic expansion. Employing this vertex {\em ansatz}, we solve the gap equation to compute dynamically generated electron mass whose dependence on the electromagnetic coupling is found to satisfy Miransky scaling law. We also investigate the gauge dependence of this dynamical mass as well as that of the critical coupling above which chiral symmetry is dynamically broken. As a litmus test of our vertex construction, both these quantities are rendered virtually gauge independent within a certain interval of values considered for the covariant gauge parameter.

\end{abstract}


\maketitle
\date{\today}

\section{Introduction}
\label{SECTION Introduction}
Graphene, the wonder material~\cite{Novoselov2005,Zhang2005}, is a physical system with immense potential for technological applications. It has driven a lot of research in both the applied and theoretical physics, not only from the point of view of condensed matter and materials sciences~\cite{CastroNeto:2007fxn,Geim2007}, but also based on the quantum field theoretic description within the domain of high energy physics~\cite{Gusynin:2007ix} and cosmology~\cite{Cortijo2007,Vozmediano_2008,VOZMEDIANO2010109}. Its remarkable properties of high electric and thermal conductivity, stiffness, flexibility and transparency have opened the door to explore a growing family of modern relativistic and relativistic-like materials in one, two and three spatial dimensions. The underlying honeycomb array of tightly packed carbon atoms and its crystallographic description in terms of two inter-imposed triangular sub-lattices provide
chiral and valley quantum labels to
the charge carrier {\em electrons}. This occurrence is responsible for Klein tunneling~\cite{Katsnelson_2006} as well as other exotic and novel phenomena~\cite{Katsnelson2006} exhibited by relativistic systems only. That makes graphene an incarnation of quantum electrodynamics (QED) in condensed matter realms. Along with quantum Hall systems~\cite{Zhang2005,Semenoff.PhysRevLett.53.2449,Haldane.PhysRevLett.61.2015,Gusynin.PhysRevLett.95.146801} and high-$T_c$ layered cuprate superconductors~\cite{DOREY1992614,PhysRevLett.87.257003,PhysRevB.66.094504,PhysRevB.66.054535,Farakos:1997qi}, graphene is also a system suitable for its description in terms of relativistic quantum field theoretical considerations, developed and refined in the domain of particle physics~\cite{Gusynin:2007ix}. This in turn allows for an exploration of otherwise inaccessible particle physics {\em phenomenology} in a more controlled and observable ambient of solid state physics. A representative example in this connection is the so-called chiral magnetic effect~\cite{Kharzeev:2007jp,Fukushima:2008xe}, which was first predicted to take place in relativistic heavy ion collisions. It involves chirality flip of quarks, caused by the chiral anomaly. It is a necessary ingredient to produce a non dissipative current as an observable effect. This effect was proposed to probe the non-tivial vacuum structure of quantum chromodynamiocs (QCD). However, it has not been observed in isobar collisions in the STAR collaboration at RHIC~\cite{STAR:2021mii}. Nevertheless the same basic idea of a physical system in which interactions drive a chirality flip of the fundamental degrees of freedom was found in ZnTe$_5$~\cite{Li:2014bha}, where a neat non-dissipative current was observed when this 3D crystal is subject to an array of adequately aligned electric and magnetic fields. After this first observation, non-dissipative currents driven by the chiral anomaly were also encountered in several other similar materials~\cite{Li2015,PhysRevX.5.031023,PhysRevB.93.121112,Arnold:2015vvs,Zhang2017}.

Some theoretical ideas have also been developed to observe a similar effect in 2D crystals such as graphene~\cite{Mizher:2018dtf,PhysRevD.102.096023}.
More recently, it has been shown that in some 2D materials, one may observe a novel quantum spin Hall phenomenon~\cite{Dudal:2021ret}. A key ingredient for the realization of these later phenomena is the description of electromagnetic and matter fields living in mixed dimensions.
Mixed dimensional theories emerge naturally in the description of 2D materials where experiments are carried out with external electromagnetic fields which permeate the whole space whereas the movement of the charge carriers remains restricted to a plane.

Two independent formulations have been proposed in literature to describe QED of photons and electrons living in different space-dimensions. One vision exploits the equivalence of a theory where electrons live in lower dimensions than photons and a Chern-Simons theory. It has been dubbed as Pseudo-QED~\cite{MARINO1993551,marino_2017,Amaral_1992,PhysRevD.90.105003}. Alternatively, a brane-world inspired scenario was developed in~\cite{PhysRevD.64.105028} to explore the traits of dynamical chiral symmetry breaking (DCSB) in the so-called Reduced QED (RQED), a nomenclature we choose to adopt in this article. The equivalence of these two visions has already been established.
It respects causality~\cite{Amaral_1992} and unitarity~\cite{PhysRevD.90.105003}. It exhibits a Coulomb static interaction in the case of graphene~\cite{marino_2017,GONZALEZ1994595}. It contains an infrared fixed point of the renormalization group as the Fermi velocity tends to the speed of light~\cite{GONZALEZ1994595,PhysRevLett.77.3589,PhysRevB.59.R2474,PhysRevB.63.134421}. Two-loop perturbative analysis has been carried out in~\cite{PhysRevD.89.065038} and later improved with renormalization group arguments~\cite{Dudal:2018pta}. Interestingly for the immediate purpose of our manuscript, DCSB has been explored within the Schwinger-Dyson~\cite{PhysRevD.87.125002} equations (SDEs) and renormalization group frameworks exploiting the duality between the gap equation in this theory and the corresponding $1/N$ leading truncation in parity preserving ordinary QED$_3$. In the latter theory, it is known that there exists a critical number of fermion families $N_c$ above which DCSB is restored. In comparison, it is observed that in the quenched version of RQED, where electron-loop contributions to the photon propagator are neglected and the photon dressing function reduces to its tree level expansion, DCSB occurs provided the electromagnetic coupling $\alpha$ exceeds a critical value $\alpha_c$. The particular values of these critical numbers depend on the gauge parameter and provide a natural motivation for the work we present and the solutions we provide in this article.
For the sake of completeness, we would like to mention that DCSB has also been considered in RQED at finite temperature and in the presence of a Chern-Simons term. Parity violating solutions to the gap equation have also been explored in~\cite{carrington.PhysRevB.99.115432} in connection with the presence of a Chern-Simons term. This term plays the role of an effective dielectric constant, hence having potential experimental realization in graphene related materials. Effects of strain have also been considered, leading to a lower value of the critical coupling required to break chiral symmetry. Finally, RQED has also been formulated in curved spaces~\cite{Caneda.PhysRevD.103.065010}. For a review discussing all this properties and applications of RQED, see \cite{Olivares:2021svj}.

Studying DCSB and its gauge invariance in RQED naturally requires its non-perturbative treatment. Fortunately, extensive amount of analogous research in QED$_4$~\cite{Curtis:1990zs,Bashir:1994az,Bashir:1995qr,Bashir:2011dp,Albino:2018ncl} and QED$_3$~\cite{Bashir:2002dz,Bashir:2004yt,Bashir:2005wt,Bashir:2009fv} provides necessary groundwork to carry out similar reliable analysis in RQED. We report the results of this continuum study through state-of-the-art truncation schemes in SDEs. Focusing on the quenched version of the theory, the sole source of our starting {\em ansatz} is the electron-photon vertex. We construct it by demanding all the key characteristics of RQED to be faithfully respected:

\begin{itemize}

    \item  Ward-Fradkin-Green-Takahashi identity (WFGTI) that connects the electron propagator with the {\em longitudinal} part of the  electron-photon vertex is satisfied non-perturbatively by construction, known as the Ball-Chiu (BC) vertex~\cite{Ball:1980ay}.

    \item To expand the transverse part of the vertex, we employ the vector basis and its coefficients in such a manner as to ensure spurious kinematic singularities are absent from our construction~\cite{Kizilersu:1995iz,Davydychev:2000rt,Bermudez:2017bpx}.
    \item In the weak coupling regime, the vertex faithfully reproduces its one-loop perturbative expansion for the asymptotic limit of momenta $k^2 \gg p^2$, just it has previously been done in QED$_4$~\cite{Curtis:1990zs,Bashir:1994az} and QED$_3$~\cite{Bashir:1999bd,Bashir:2000rv,Bashir:2011vg}.

    \item The standard model of particle physics tells us of the intimate connection between its renormalizability and gauge invariance. In the same spirit, we require our vertex {\em ansatz} to guarantee the multiplicative renormalizability (MR) of the massless electron propagator in its leading logarithmic expansion.

    \item We also require our vertex to satisfy the discrete symmetries of parity, time reversal and charge conjugation.

\end{itemize}

Based upon the above-mentioned constraints, we are able to achieve {\em nearly} gauge independent Euclidean mass and critical coupling $\alpha_c$ where the DCSB solution bifurcates away from the chirally symmetric one. We believe that obtaining gauge independent DCSB holds the promise to study observable effects in the $2D$ materials described by RQED in a reliable manner through continuum SDEs.

The article has been organized as follows: Sect.~{\ref{SECTION RQED foundations}} begins
with a brief introduction to the mathematical foundations of RQED. In Sect.~{\ref{SECTION Vertex decomposition}}, we provide necessary preliminaries on the vertex decomposition and its general features. In Sect.~{\ref{SECTION One-loop vertex}}, we construct a family of {\em Ans$\ddot{a}$tze} for the transverse vertex in perhaps the most economical yet efficient manner by resorting to the constraints of MR and its explicit form in the so-called asymptotic limit at the one-loop level.
In Sect.~{\ref{SECTION Gap equation}}, we set up the gap equation and engage in a detailed discussion on the photon propagator in RQED and the appropriate use of the WFGTI in order to ensure the MR of the massless electron propagator.  Sect.~{\ref{SECTION Solution gap equation}}
provides solution of the gap equation, first in the perturbative realm and then the non-perturbative DCSB solution in terms of the critical coupling $\alpha_c$ above which the massive solution bifurcates away from the perturbative massless one. We primarily focus on obtaining gauge independent DCSB.  Sect.~{\ref{SECTION Conclusions and Perspectives}} contains conclusions and offers perspectives for future work.

\section{RQED foundations}
\label{SECTION RQED foundations}

In order to describe electrons, restricted to move in a dimensionally reduced space-time, coupled to photons free to propagate through the bulk space-time, one initially begins with the well-known QED Lagrangian:
\begin{eqnarray}
  \mathcal{L}_{\text{QED}} &=& \bar{\psi} \left( i \gamma^{\mu} \partial_{\mu} - m_0 \right) \psi + j_\mu A^\mu\nonumber \\
  && -\frac{1}{4}F_{\mu\nu}F^{\mu\nu} - \frac{1}{2 \xi} \left( \partial_{\mu} A^{\mu} \right)^2 \,,
\end{eqnarray}
where $\psi$ and $A^{\mu}$ are electron and photon fields, respectively, coupled to each other through the electromagnetic current $j_{\mu}$, where $\mu = 0,1,2,3$. Furthermore, the 4-dimensional Dirac matrices $\gamma^{\mu}$ satisfy the anti-commutation relation $\left\{ \gamma^{\mu} , \gamma^{\nu} \right\} = 2 g^{\mu\nu}$, with the commonly used convention $g^{\mu\nu} = (+,-,-,-)$ for the Minkowski space metric tensor. Additionally, $m_0$ is the bare electron mass, $\xi$ is the covariant gauge parameter and $F_{\mu\nu} = \partial_{\mu} A_{\nu} - \partial_{\nu} A_{\mu}$ is the usual electromagnetic field tensor. The corresponding action $\mathcal{S}_{\text{QED}} = \int \dd^4 x \mathcal{L}_{\text{QED}}$ can be conveniently expressed as
\begin{eqnarray}
  \mathcal{S}_{\text{QED}} = \mathcal{S}_{\bar{\psi}\psi}^{(3)} + \frac{1}{2} \int \dd^4x \left[ j^{\mu} \hat{\Delta}_{\mu \nu} j^{\nu} - A^{\mu} \hat{\Delta}_{\mu \nu}^{-1} A^{\nu} \right] \,,
  \label{QED action}
\end{eqnarray}
where the kinetic term for electrons, constrained to move in a 3-dimensional space-time reads:
\begin{eqnarray}
\mathcal{S}_{\bar{\psi}\psi}^{(3)} = \int \dd^3 x \, \bar{\psi} \left( i \gamma^{\mu} \partial_{\mu} - m_0 \right) \psi \,,
\end{eqnarray}
with $\mu = 0,1,2$. Moreover, the differential operator $\hat{\Delta}_{\mu \nu}$, namely the photon propagator in coordinate space, can be cast in terms of its momentum space counterpart by means of a Fourier transformation
\begin{eqnarray}
\hat{\Delta}_{\mu\nu} &=& - \hspace{-1mm} \int \hspace{-1.5mm} \frac{\dd^4 q}{\left( 2\pi\right)^4} e^{ -i q \cdot x } \frac{1}{q^2} \left[ g_{\mu\nu} - \left(1-\xi\right) \frac{q_{\mu} q_{\nu}}{q^2} \right] \,,
\label{Photon Prop Operator in 4D}
\end{eqnarray}
satisfying $\hat{\Delta}_{\mu \alpha}^{-1} \hat{\Delta}^{\alpha \nu} = g_{\mu}^{\nu}$ with its corresponding inverse propagator. Such a Green function accounts for a gauge field propagating through the whole 4-dimensional space-time with $\mu, \nu = 0,1,2,3$.

To account for a mixed-dimension system described by RQED with electrons restricted to move on a plane perpendicular to the $x_3$-axis~\cite{MARINO1993551,marino_2017,Amaral_1992,PhysRevD.90.105003,PhysRevD.64.105028,PhysRevD.86.025005}, the electromagnetic current takes the form
\begin{eqnarray}
  j^{\mu} &=& \left\{\begin{array}{cc}
              -i e \bar \psi \gamma^\mu \psi \delta(x_3) &  ~\text{for}~\mu=0,1,2\,, \\
              0& ~\text{for}~\mu=3\,.
            \end{array}\right.
\end{eqnarray}
Therefore, only the indices $\mu=0,1,2$ contribute to the term $ j^{\mu} \hat{\Delta}_{\mu \nu} j^{\nu}$ in
Eq.~(\ref{QED action}). The component $\mu=3$ can thus be integrated out
in Eq.~(\ref{Photon Prop Operator in 4D}), leading to \cite{Dudal:2018mms}
\begin{eqnarray}
\hat{\Delta}_{\mu\nu} &=& \hspace{-1.5mm} \int \hspace{-1.5mm} \frac{\dd^3 q}{\left( 2\pi\right)^3} e^{ -i q \cdot x } \frac{1}{2 \sqrt{-q^2}} \left[ g_{\mu\nu} - \left(1-\xi\right) \frac{q_{\mu} q_{\nu}}{2 q^2} \right] \,.
\label{Photon Prop Operator in 3D}
\end{eqnarray}
It entails a non-local differential operator. This propagator can be obtained from the effective action for RQED (redefining $\xi = 2\zeta - 1$)
\begin{eqnarray}
  \mathcal{S}_{\text{RQED}} &=& \mathcal{S}_{\bar{\psi}\psi}^{(3)} + \int \dd^3 x \, j_{\mu} A^{\mu} \nonumber \\
  && \hspace {-12mm}+ \hspace{-1.5mm} \int \hspace{-1.5mm} \dd^3x \left[ \frac{1}{2}F_{\mu\nu} \frac{1}{\sqrt{- \partial^2}} F^{\mu\nu} + \frac{1}{\zeta} \partial_{\mu} A^{\mu} \frac{1}{\sqrt{- \partial^2}} \partial_{\nu} A^{\nu} \right] .
  \label{RQED action}
\end{eqnarray}
From now on, we work in the Euclidean space defined by the metric tensor $\delta_{\mu\nu} = (+,+,+)$ for $\mu,\nu = 4,1,2$. In this space, the bare photon propagator takes the form (\textit{c.f.}
Eq.~(\ref{Photon Prop Operator in 3D}))
\begin{eqnarray}
\Delta_{\mu\nu}^{(0)} (q) = \frac{1}{2 q} \left[ \delta_{\mu\nu} - \left(1-\xi\right) \frac{q_{\mu} q_{\nu}}{2 q^2} \right] \,,
\label{bare photon prop}
\end{eqnarray}
where we have defined $q \equiv \sqrt{-q^2}$. Note that this propagator has a softer infrared behavior than the photon propagator in QED$_4$ and QED$_3$. Several groups differ in conventions by the global factor of 1/2. When written in terms of the variable $\zeta$, this propagator can be separated into a familiar longitudinal and a transverse component (to $q_{\mu}$):
\begin{eqnarray}
\Delta_{\mu\nu}^{(0)} (q) = \frac{1}{2 q} \left[ \delta_{\mu\nu} - \frac{q_{\mu} q_{\nu}}{q^2} \right] + \zeta \frac{q_{\mu} q_{\nu}}{2 q^3} \,,
\label{bare photon prop zeta}
\end{eqnarray}
However, it must be emphasized that the gauge fixing parameter $\zeta$ in RQED is different from that in QED, $\xi$, due to dimensional reduction. In the above expression for the photon propagator, Eq.~(\ref{bare photon prop zeta}), the $\zeta$-independent term defines the transverse propagator which is the only component that gets quantum corrections. This is the reason for using such a decomposition in other works. In contrast, the $\xi$-independent term in Eq.~(\ref{bare photon prop}) does not define a transverse propagator. Therefore, quantum corrections affect both $\xi$-dependent and $\xi$-independent components. In the present work, we restrict ourselves to work in the quenched approximation. As there are no fermion loops present, there are no quantum corrections to the bare photon propagator.
Therefore, both the expression, Eqs.~(\ref{bare photon prop},\ref{bare photon prop zeta}) are equally suitable: we choose that of Eq.~(\ref{bare photon prop}).

Note that the electron-photon vertex plays a vital role in computing the non-perturbative solution of the electron propagator through its gap equation. Therefore, we address this three-point Green function at length in the following section.

\section{The Vertex: Generalities}
\label{SECTION Vertex decomposition}

\begin{figure}[!ht]
    \centering
    \includegraphics[scale=.5]{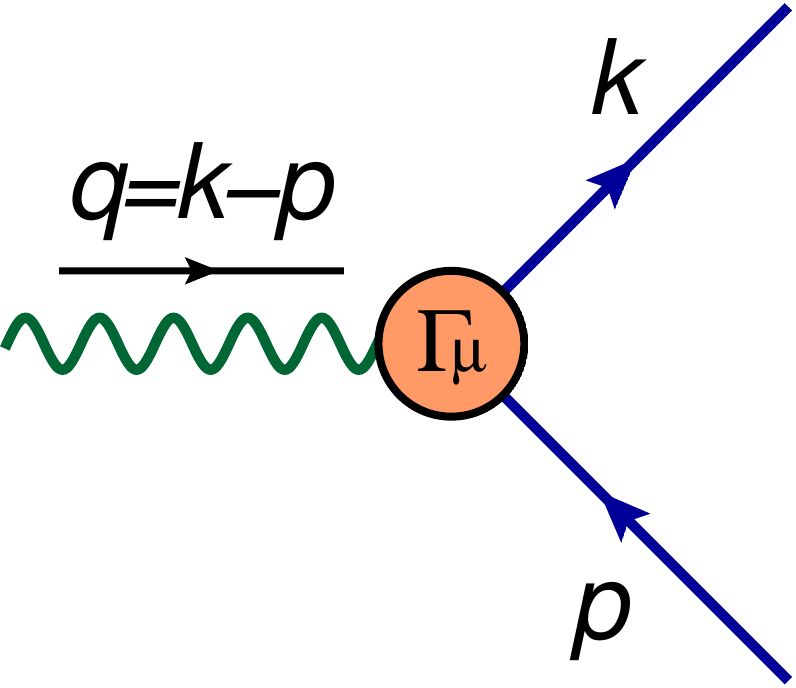}
    \caption{Diagrammatic representation of the full electron-photon vertex
$\Gamma_{\mu}(k,p)$, with momentum flow indicated.}
    \label{Vertice3PuntosLuis}
\end{figure}

In its general decomposition, the three-point electron-photon vertex can be written in terms of twelve independent spin structures. For the kinematical configuration of Fig.~\ref{Vertice3PuntosLuis}, the WFGTI
associated with this vertex takes the form
\begin{equation}
i q_{\mu}\Gamma_{\mu}(k,p)=S^{-1}(k)-S^{-1}(p) \,, \label{WGTI for the 3-point vertex}
\end{equation}
where $q=k-p$. This identity allows us to split the vertex as
a sum of \textit{longitudinal} and \textit{transverse} components,
as suggested by Ball and Chiu~\cite{Ball:1980ay}:
\begin{equation}
\Gamma_{\mu}(k,p)=\Gamma^{L}_{\mu}(k,p)+\Gamma^{T}_{\mu}(k,p) \,. \label{Ball-Chiu vertex decomposition}
\end{equation}
The longitudinal part $\Gamma^{L}_{\mu}(k,p)$ alone satisfies the WFGTI (\ref{WGTI for the 3-point vertex}), and consumes four of the twelve independent spin structures (one of them is zero in QED), so that,~\cite{Ball:1980ay}:
\begin{eqnarray}
\hspace{-6mm}\Gamma^{L}_{\mu}(k,p)= \lambda_1(k,p) \gamma_{\mu} +
\lambda_2(k,p) t_{\mu} \gamma \cdot t
 - i  \lambda_3(k,p) t_{\mu} \,, \label{Longitudinal vertex decomposition}
\end{eqnarray}
with $t=k+p$, and
\begin{eqnarray}
\lambda_1(k,p) &=& \frac{1}{2} \left[
\frac{1}{F(k^{2},\Lambda^{2})} +
\frac{1}{F(p^{2},\Lambda^{2})} \right] \,, \nonumber \\
\lambda_2(k,p) &=&  \frac{1}{2}  \left[ \frac{1}{F(k^{2},\Lambda^{2})} -
\frac{1}{F(p^{2},\Lambda^{2})} \right] \frac{1}{k^{2}-p^{2}} \,, \nonumber \\
\lambda_3(k,p) &=& \left[ \frac{ {\cal{M}}(k^{2},\Lambda^{2})
}{F(k^{2},\Lambda^{2})} - \frac{ {\cal{M}}(p^{2},\Lambda^{2})
}{F(p^{2},\Lambda^{2})} \right] \frac{1}{k^{2}-p^{2}} \,, \label{longitudinal coefficients definitions}
\end{eqnarray}
where $\Lambda$ is an ultraviolet (UV) cut-off regulator. Note that ${\cal{M}}(k^{2},\Lambda^{2})$
and $F(k^{2},\Lambda^{2})$ are the mass function and the wave function
renormalization, respectively, related to the electron propagator $S(k)$
through
\begin{equation}
S(k) = \frac{F(k^{2},\Lambda^{2})}{i \gamma \cdot k +
{\cal{M}}(k^{2},\Lambda^{2})} \,, \label{fermion propagator definition}
\end{equation}
with $F(k^{2},\Lambda^{2})=1$ and
${\cal{M}}(k^{2},\Lambda^{2})=m_0$ at the tree level.

The transverse part $\Gamma^{T}_{\mu}(k,p)$ of the vertex
decomposition (\ref{Ball-Chiu vertex decomposition}), which
remains undetermined by the WFGTI, is naturally constrained by
\begin{equation}
 q_{\mu}\Gamma^{T}_{\mu}(k,p)=0 \,. \label{transverse part definition}
\end{equation}
In general, the ultraviolet finite transverse vertex can be
expanded out in terms of eight basis vector structures, and their
corresponding scalar form factors $\tau_{i}(k,p)$~\cite{Ball:1980ay}:
\begin{equation}
\Gamma^{T}_{\mu}(k,p) = \sum_{i=1}^{8} \tau_{i}(k,p)
T^{i}_{\mu}(k,p)\,. \label{transverse vertex structure}
\end{equation}
For the kinematical configuration of Fig.~\ref{Vertice3PuntosLuis}, we define
\begin{eqnarray}
T^{1}_{\mu}(k,p) &=& i \left[ p_{\mu} (k \cdot q) -k_{\mu} (p
\cdot q) \right] \,, \nonumber \\
T^{2}_{\mu}(k,p) &=& \left[ p_{\mu} (k \cdot q) -k_{\mu} (p \cdot
q) \right] \left( \gamma \cdot t \right) \,, \nonumber \\
T^{3}_{\mu}(k,p) &=& q^{2} \gamma_{\mu} - q_{\mu} \left( \gamma \cdot q \right) \,, \nonumber \\
T^{4}_{\mu}(k,p) &=& i q^{2} \left[ \gamma_{\mu} \left( \gamma \cdot t \right) -
t_{\mu} \right] + 2 q_{\mu} p_{\nu} k_{\rho} \sigma_{\nu \rho}  \,, \nonumber \\
T^{5}_{\mu}(k,p) &=& \sigma_{\mu \nu} q_{\nu}  \,, \nonumber \\
T^{6}_{\mu}(k,p) &=& - \gamma_{\mu} \left( k^{2}-p^{2} \right) +
t_{\mu} \left( \gamma \cdot q \right) \,, \nonumber \\
T^{7}_{\mu}(k,p) &=&  \frac{i}{2} (k^{2}-p^{2}) \left[
\gamma_{\mu} \left( \gamma \cdot t \right) - t_{\mu} \right] + t_{\mu} p_{\nu}
k_{\rho} \sigma_{\nu  \rho}   \,, \nonumber \\
T^{8}_{\mu}(k,p) &=& -i \gamma_{\mu} p_{\nu} k_{\rho} \sigma_{\nu
 \rho} - p_{\mu}  \left( \gamma \cdot k \right) + k_{\mu} \left( \gamma \cdot p \right) \,,
\label{transverse basis definition}
\end{eqnarray}
with
\begin{equation}
\sigma_{\nu \rho} = \frac{i}{2} \left[
\gamma_{\nu},\gamma_{\rho} \right] \,. \label{sigma definition}
\end{equation}
This basis is not exactly the one adopted in~\cite{Ball:1980ay}. We
choose to work with a modification of this initial basis which was put forward
in~\cite{Kizilersu:1995iz} and later employed in~\cite{Davydychev:2000rt} as well.
This modified choice of the basis vectors ensures all transverse form factors of the vertex are independent of any kinematic singularities in one-loop perturbation theory in an arbitrary covariant gauge.

As stated earlier in Sect. \ref{SECTION Introduction}, any
\textit{ansatz} for the full vertex must have the same
transformation properties as the bare vertex under charge
conjugation operation. This requires all the $\tau_i$s in (\ref{transverse
vertex structure}) to be symmetric under the interchange $k
\leftrightarrow p$, except $\tau_{4}$ and $\tau_{6}$, which are
odd:
\begin{eqnarray}
&& \hspace{-.5cm} \tau_{i}(k,p)= \tau_{i}(p,k) \,, \hspace{.625cm}
i=1,2,3,5,7,8, \label{tau symmetric properties} \\
&& \hspace{-.5cm} \tau_{i}(k,p) = -\tau_{i}(p,k) \,, \quad i=4,6.
\label{tau antisymmetric properties}
\end{eqnarray}
From Eq.~(\ref{longitudinal coefficients definitions}), it is
obvious that $\lambda_1(k,p)$, $\lambda_2(k,p)$ and
$\lambda_3(k,p)$ are symmetric under $k \leftrightarrow p$, as
they should be, in order to preserve the correct transformation
properties under charge conjugation operation for the full vertex.

In the following section, we construct the transverse vertex, putting forward an {\em ansatz} which complies with the key requirements expected of it as detailed in Sect. I. It is also effective, economical and yields practically gauge invariant DCSB as we see in subsequent sections.



\section{The transverse vertex}
\label{SECTION One-loop vertex}

The transverse vertex is completely determined once the form factors in Eqs.~(\ref{transverse vertex structure},\ref{transverse basis definition}) are
known. We now proceed to construct an {\em ansatz} for it.

\subsection{A general {\em ansatz} for the transverse vertex}

We start out by recalling a fairly general {\em ansatz} which was first proposed in~\cite{Bashir:2011dp}, generally referred to as the BB-vertex, and later successfully employed in several related works, e.g.,~\cite{Bermudez:2017bpx,Albino:2018ncl,Albino:2021rvj}
\begin{eqnarray}
\tau_{1}(k,p) &=& \frac{a_{1}}{(k^{2}+p^{2})} \; \lambda_3(k,p) \,, \label{Rocio ansatz tau1} \\
\tau_{2}(k,p) &=& \frac{2 a_{2}}{(k^{2}+p^{2})} \; \lambda_2(k,p) \,, \label{Rocio ansatz tau2} \\
\tau_{3}(k,p) &=& 2 a_{3} \; \lambda_2(k,p) \,, \label{Rocio ansatz tau3} \\
\tau_{4}(k,p) &=& \frac{a_{4} (k^2-p^2) }{4 k^2 p^2} \; \lambda_3(k,p) \,, \label{Rocio ansatz tau4} \\
\tau_{5}(k,p) &=& - a_{5} \; \lambda_3(k,p) \,, \label{Rocio ansatz tau5} \\
\tau_{6}(k,p) &=& - \frac{2 a_{6} (k^2+p^2)}{(k^2-p^2)} \; \lambda_2(k,p)
\,, \label{Rocio ansatz tau6} \\
\tau_{7}(k,p) &=& -\left[  \frac{ a_{4} q^{2} }{ 2 k^{2}
p^{2} } +  \frac{a_{7}}{ k^{2} + p^{2} } \right] \; \lambda_3(k,p) \,, \label{Rocio ansatz tau7} \\
\tau_{8}(k,p) &=& 2 a_{8} \; \lambda_2(k,p) \,. \label{Rocio ansatz tau8}
\end{eqnarray}
Before we proceed any further, we summarize the following important points:
\begin{itemize}
    \item All eight transverse form factors are taken into consideration which implies the generality of this {\em ansatz}.
    \item All form factors are proportional to the same structures which appear in the {\em longitudinal} vertex, namely $\lambda_2(k,p)$ and $\lambda_3(k,p)$.
    \item The form factor $\tau_6(k,p)$ has a kinematic singularity for $k^2=p^2$. For the QED$_4$ gap equation, we can employ this form factor with impunity because it cancels with the corresponding term proportional to $\lambda_2(k,p)$ coming from the {\em longitudinal} vertex.
    However, for RQED, this cancellation does not take place and we need to modify this form factor to avoid the kinematic singularity. We propose an economical and effective modification:
\begin{eqnarray}
\tau_{6}(k,p) &=& a_{6} \frac{(k^2-p^2)}{(k^2+p^2)} \; \lambda_2(k,p) \;.
\end{eqnarray}
\end{itemize}
We now try to constrain the coefficients $a_i$ through the one-loop expression for the {\em transverse} vertex. We may call this as the modified BB-vertex for RQED.

\subsection{One-Loop electron-photon vertex}
In perturbation theory, the one-loop order expansion of the electron-photon vertex is
\begin{eqnarray}
&& \hspace{-4mm} \Gamma_{\mu}^{(1)} (k,p) = \gamma_{\mu} \nonumber \\
&& - \frac{\alpha}{2\pi^2} \int{ d^3 \omega \gamma_{\alpha} S(p-\omega) \gamma_{\mu} S(k-\omega) \gamma_{\beta} \Delta_{\alpha \beta}(\omega) } \,.
\end{eqnarray}
In analogy with~\cite{Curtis:1991fb,Bashir:1997qt,Bashir:2011vg}, we have computed the one-loop corrections to the electron-photon vertex in the asymptotic limit, defined as the perturbative expansion with $p^2 \gg k^2 \gg m^2_0$. The leading logarithmic term of the transverse vertex at this level of approximation reads:
\begin{eqnarray}
\Gamma_{\mu}^{T} (k,p) & \overset{p^2\gg k^2}{=} & - \frac{\alpha }{8 \pi p^2} \left( \xi - \frac{1}{3} \right)\log \left( \frac{p^2}{k^2} \right) T_{\mu}^{asy} \,, \label{asymptotic vertex}
\end{eqnarray}
with
\begin{eqnarray}
T_{\mu}^{asy} = p^2 \gamma_{\mu} - p_{\mu} \gamma \cdot p \,.
\end{eqnarray}
Hence, in this limit, the transverse vertex can be expressed as (\textit{cf.} Eq.~(\ref{Lead Log Result}))
\begin{eqnarray}
\Gamma_{\mu}^{T} (k,p) & \overset{p^2\gg k^2}{=} & \frac{1}{2} \left[ \frac{1}{F(k^2)} - \frac{1}{F(p^2)} \right] \frac{T_{\mu}^{asy}}{k^2 - p^2} \,. \label{CP asymptotic vertex}
\end{eqnarray}
On the other hand, from Eqs.~(\ref{transverse vertex structure},\ref{transverse basis definition}), it is straightforward to see that the leading structure of the transverse vertex in this limit acquires the following form:
\begin{eqnarray}
\Gamma_{\mu}^{T} (k,p)  & \overset{p^2\gg k^2}{=} &
\left( \tau_3 + \tau_6 \right) T_{\mu}^{asy} \,,
\label{Asymptotic vertex}
\end{eqnarray}
where $\tau_{3,6} \equiv \tau_{3,6}(k,p)$. Moreover, we have used the fact that, in the asymptotic expansion, the dominant contributions to the transverse vertex come from $T_{\mu}^{3}$ and $T_{\mu}^{6}$ which simplify to
\begin{eqnarray}
T_{\mu}^{3 \, asy} = T_{\mu}^{6 \, asy} \equiv T_{\mu}^{asy} \, .\label{Asymptotic basis}
\end{eqnarray}
As the simplest construction of the transverse vertex, we can single out $\tau_3(k,p)$ and $\tau_6(k,p)$ which correctly provide this desired limit. Writing out these form factors explicitly,
\begin{eqnarray}
\tau_3(k,p) &=& a_3 \left[ \frac{1}{F(k^2)} - \frac{1}{F(p^2)} \right] \frac{1}{k^2 - p^2} \,, \label{Tau 3 Asymptotic} \\
\tau_6(k,p) &=& a_6 \left[ \frac{1}{F(k^2)} - \frac{1}{F(p^2)} \right] \frac{1}{k^2 + p^2} \,. \label{Tau 6 Asymptotic}
\end{eqnarray}
It is thus straightforward to see from Eqs.~(\ref{CP asymptotic vertex},\ref{Asymptotic vertex}) that the one-loop behaviour of the vertex in the asymptotic limit requires
\begin{eqnarray}
a_3 - a_6 = 1/2 \,.
\label{PT restricion on a36}
\end{eqnarray}
Our aim is to compute non-perturbative solutions of the gap equation using various vertex {\em ans\"atze}. Our preferred choice is the electron-photon vertex consonant with the WFGTI and perturbation theory. It consists of the longitudinal BC-vertex, Eq.~(\ref{Longitudinal vertex decomposition}), and the transverse component constructed in terms of the vector structures $T_{\mu}^3$ and $T_{\mu}^6$ alone.
For the sake of completeness, we also compute and depict some results obtained by employing the bare vertex, the central BC-vertex and the full BC-vertex.

The starting point for such an endeavor is naturally the gap equation for the electron propagator. This is what we proceed to set up and study in the next section.


\section{Setting up the gap equation}
\label{SECTION Gap equation}

\begin{figure}[!ht]
    \centering
    \includegraphics[scale=.5]{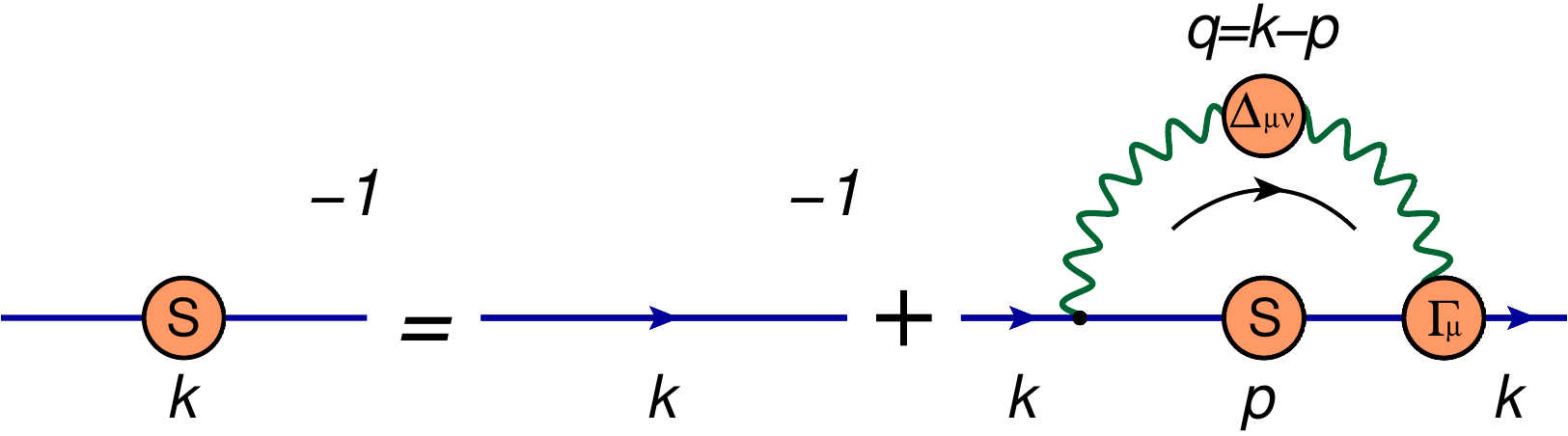}
    \caption{The gap equation for the electron propagator. The color-filled
    blobs labelled with $S$, $\Delta_{\mu\nu}$ and
    $\Gamma_{\mu}$ stand for the fully-dressed electron and photon
    propagators, and the three-point vertex, respectively.}
    \label{PropagatorSDEQED}
\end{figure}

The SDE for the electron propagator, also known as the electron
\textit{gap equation}, is diagrammatically depicted in
Fig.~\ref{PropagatorSDEQED}.

Mathematically, the gap equation is written as:
\begin{eqnarray}
S^{-1}(k) &=& S_{0}^{-1}(k) + \frac{\alpha}{2 \pi^{2}} \int_{E}{
d^{3}p \, \gamma_{\nu} S(p) \Gamma_{\mu}(k,p) \Delta_{\mu\nu} (q)
} \,, \nonumber \\
 \label{gap equation}
\end{eqnarray}
where the subscript $E$ indicates that the integral is performed in the Euclidean space, the subscript ``0" denotes the tree level of the corresponding propagator and $\alpha=e^{2}/4 \pi$ is the electromagnetic coupling. We have already provided a detailed discussion on the full electron-photon vertex $\Gamma_{\mu}(k,p)$. A similar analysis of the photon propagator
$\Delta_{\mu\nu} (q)$ is in place now.

\subsection{The Photon Propagator}

For the sake of an appropriate implementation of the WFGTI in the gap equation, let us split the photon propagator of Eq.~(\ref{bare photon prop}) in a longitudinal piece
\begin{eqnarray}
\Delta_{\mu\nu}^{L} (q) &=& \varrho \, \kappa \frac{q_{\mu}
q_{\nu}}{q^{3}}\,, \label{Photon Propagator Longitudinal}
\end{eqnarray}
and a remaining \textit{non-longitudinal} component
\begin{eqnarray}
\Delta_{\mu\nu}^{NL} (q) &=& \, \frac{\varrho}{q} \left[ \delta_{\mu \nu} -  \chi
\frac{q_{\mu} q_{\nu}}{q^{2}} \right]\,, \label{Photon Propagator Non-Longitudinal}
\end{eqnarray}
expressed in Euclidean space such that
\begin{eqnarray}
\Delta_{\mu\nu} (q) &=& \Delta_{\mu\nu}^{L}(q) + \Delta_{\mu\nu}^{NL}(q) \,, \label{Photon propagator definition}
\end{eqnarray}
where $\chi \equiv 1 - \epsilon_{e}$, $\kappa \equiv \chi \xi$, $\varrho = \Gamma\left[ \chi \right] / \left( 4\pi \right)^{\epsilon_{e}}$.
It is important to notice that RQED ($\epsilon_{e}=1/2$) yields $\chi = \varrho = 1/2$ and $\kappa = \xi/2$. We have opted for this notation of arbitrary dimensions in alignment with the one adopted in~\cite{PhysRevD.86.025005}.

This separation of the photon propagator, Eqs.~(\ref{Photon Propagator Longitudinal},\ref{Photon Propagator Non-Longitudinal}),
plays an important role in QED where it has been demonstrated that it is imperative to apply the WFGTI to the divergence term $q_{\mu} \Gamma_{\mu}$ which arises from the contraction of the vertex with the longitudinal (gauge dependent) contribution of the photon propagator in the gap equation. This recipe ensures the absence of spurious terms which are a characteristic of the UV cut-off regularization scheme due to the fact that it breaks translational invariance and can violate gauge invariance if not used carefully~\cite{Dong:1994jr,Bashir:1994az}. Providing a vertex that satisfies the WFGTI and employing  this identity in the gap equation only for the appropriate part of the photon propagator are crucial to guarantee the local gauge covariance of the electron propagator.
In the quenched approximation, the electron wave function renormalization for QED$_4$ has been shown to be explicitly  renormalizable to all orders in perturbation theory provided an {\em ansatz} for the vertex in close analogy with full QED$_4$ is constructed~\cite{Curtis:1990zs}. In the leading logarithmic approximation, the electron wave function renormalization exhibits a power law behaviour:
\begin{eqnarray}
F(k^{2}) &=& 1 + \sum_{n=1}^{\infty} \frac{\beta^{n}}{n!} \log ^{n} \left( \frac{k^{2}}{\Lambda^{2}} \right) = \left( \frac{k^{2}}{\Lambda^{2}} \right)^{\beta}  \,, \label{F Power Law}
\end{eqnarray}
where $\beta = \beta(\alpha)$ is an unknown coefficient that can be computed order by order in perturbation theory. In the leading logarithmic approximation of QED$_4$, it is known that $\beta = \alpha \xi / (4 \pi)$. For quenched RQED we demonstrate that $\beta = \alpha (\xi -1/3) / (4 \pi)$.  

As mentioned earlier, careless implementation of an ultraviolet cut-off regularization breaks gauge invariance and it is manifest in the appearance of spurious terms even in one-loop calculations. An appropriate use and manipulation of the WFGTI in the gap equation is key to getting rid of such spurious terms. For this purpose, it is worth noting that in quenched RQED we can redefine the longitudinal and non-longitudinal components of the photon propagator by shifting $\chi$ and $\kappa$ with an arbitrary factor $\eta$ such that the complete photon propagator, Eq.~(\ref{Photon propagator definition}), remains the same:
\begin{eqnarray}
\chi & \rightarrow & \chi - \eta \,, \label{Shift Chi} \\
\kappa & \rightarrow & \kappa - \eta \,. \label{Shift Kappa}
\end{eqnarray}
Bear in mind that the photon propagator and consequently the electron SDE, Eq.~(\ref{gap equation}), remain invariant under the shifts prescribed by Eqs.~(\ref{Shift Chi},\ref{Shift Kappa}),

We apply the WFGTI, Eq.~(\ref{WGTI for the 3-point vertex}), on the longitudinal photon propagator term in the gap equation without loss of generality, leading to
\begin{eqnarray}
\hspace{-.5cm} \int_{E}{ d^{3}p \, \gamma_{\nu} S(p) \Gamma_{\mu}(k,p) \Delta^{L}_{\mu\nu} (q) } &=&  \varrho \, \kappa \int_{E}{ \frac{d^{3}q}{q^{3}} \gamma \cdot q } \nonumber \\
&& \hspace{-2cm} - i \varrho \, \kappa \int_{E}{ \frac{d^{3}p}{q^{3}} \gamma \cdot q \, S(p) S^{-1}(k) } \,,
\end{eqnarray}
where we have shifted $d^3p \rightarrow d^3q$ in the first term on the right-hand side of the above equation in order to show that such a term vanishes in a translationally invariant theory as it is an odd integral.
Moreover, for the term $\Gamma_{\mu}(k,p) \Delta^{NL}_{\mu\nu} (q)$ in Eq.~(\ref{gap equation}) we use the explicit form of the vertex defined through Eqs.~(\ref{Ball-Chiu vertex decomposition}-\ref{longitudinal coefficients definitions},\ref{transverse vertex structure},\ref{transverse basis definition}). After bringing out the subtle role played by the photon propagator in RQED, we can now focus on the mathematical and technical details of the gap equation itself.

\subsection{The gap equation}

We can project out two coupled, integral equations
for ${\cal{M}}$ and $F$ from the matrix gap equation. In an arbitrary gauge, these equations can respectively be written as:
\begin{eqnarray}
\frac{ {\cal{M}} (k^2) }{ F(k^2) } &=& m_0 + \frac{\alpha \varrho \kappa}{2 \pi^2} \int_{E} \frac{d^3 p}{q^3} \frac{F(p^2)}{ p^2 + {\cal{M}}^2
(p^2) } \frac{1}{F(k^2)} \nonumber \\
&& \hspace{2cm} \times \Big\{ {\cal{M}} (p^2) \, q \cdot k -
{\cal{M}} (k^2) \, q \cdot p \Big\} \nonumber \\
&& \hspace{3mm} +\frac{\alpha \varrho}{2 \pi^2} \int_{E} d^3 p \, \frac{F(p^2)}{ p^2 +
{\cal{M}}^2 (p^2) } \,
G_{{\cal{M}}} (k,p) \,,
 \label{ProyM}
\end{eqnarray}
\begin{eqnarray}
\frac{ 1 }{ F(k^2) } &=& 1 - \frac{\alpha \varrho \kappa}{2 \pi^2} \int_{E}
\frac{d^3 p}{q^3} \frac{F(p^2)}{ p^2 + {\cal{M}}^2 (p^2) }
\frac{1}{F(k^2)} \nonumber \\
&& \hspace{2cm} \times \Big\{ \, q \cdot p + {\cal{M}} (k^2)
{\cal{M}} (p^2) \, \frac{q \cdot k}{k^2} \Big\} \nonumber \\
&& \hspace{3mm} + \frac{\alpha \varrho}{2 \pi^2} \int_{E} \frac{d^3 p}{k^2}
\frac{F(p^2)}{ p^2 + {\cal{M}}^2 (p^2) } \, G_F (k,p) \,, \label{ProyF}
\end{eqnarray}
where we have adopted the notation $F(k^{2}) \equiv F(k^{2},\Lambda^{2})$ and the same for ${\cal{M}}$. Moreover, the electron-photon vertex form factors contribute to the gap equation via the scalar functions $G_{{\cal M}}$ and $G_{F}$:
\begin{eqnarray}
q \, G_{{\cal M}} (k,p) &=& (3-\chi) {\cal M}(p^2) \lambda_1 \nonumber \\
&& \hspace{-7mm} + \left[ t^2 - \frac{(k^2-p^2)^2}{q^2}\chi \right] {\cal M}(p^2) \lambda_2 \nonumber \\
&& \hspace{-7mm} - \left[ t\cdot p - \frac{(q\cdot p) (k^2-p^2)}{q^2}\chi \right] \lambda_3 \nonumber \\
&& \hspace{-7mm} + \nabla(k,p) \tau_1 + 2 \nabla(k,p) {\cal M}(p^2) \tau_2 \nonumber \\
&& \hspace{-7mm} + 2 q^2 {\cal M}(p^2) \tau_3 \nonumber \\
&& \hspace{-7mm} - 2 \left[ (k^2-p^2) (q\cdot p) + \nabla(k,p) \right] \tau_4 \nonumber \\
&& \hspace{-7mm} - 2 (q\cdot p) \, \tau_5 - 2(k^2-p^2) {\cal M}(p^2) \tau_6 \nonumber \\
&& \hspace{-7mm} - \left[ (k^2-p^2) (t\cdot p) - \nabla(k,p) \right] \tau_7 \,,
\end{eqnarray}

\begin{eqnarray}
q \,G_F (k,p) &=& \left[ (1-3\chi) k\cdot p + 2\chi \, u(k,p) \right] \lambda_1 \nonumber \\
&& \hspace{-7mm} - \left[ (k\cdot p)t^2 + 2 \nabla(k,p) - \frac{(k\cdot p) (k^2-p^2)^2}{q^2}\chi \right] \lambda_2 \nonumber \\
&& \hspace{-7mm} - \left[ t\cdot k - \frac{(q\cdot k) (k^2-p^2)}{q^2}\chi \right] {\cal M}(p^2) \lambda_3 \nonumber \\
&& \hspace{-7mm} + \nabla(k,p) {\cal M}(p^2) \tau_1 - (k^2+p^2) \nabla(k,p) \tau_2 \nonumber \\
&& \hspace{-7mm} + 2 (q\cdot k)(q\cdot p) \tau_3 \nonumber \\
&& \hspace{-7mm} + 2 \left[ q^2 (t\cdot k) - \nabla(k,p) \right] {\cal M}(p^2) \tau_4 \nonumber \\
&& \hspace{-7mm} + 2 (q\cdot k) {\cal M}(p^2) \tau_5 - 2 (k^2-p^2) (k\cdot p) \, \tau_6 \nonumber \\
&& \hspace{-7mm} + \left[ (k^2-p^2) (t\cdot k) + \nabla(k,p) \right] {\cal M}(p^2) \tau_7 \nonumber \\
&& \hspace{-7mm} + \nabla(k,p) \tau_8 \,,
\end{eqnarray}
where we have used the simplifying notation $\lambda_i \equiv \lambda_i(k,p)$ and $\tau_i \equiv \tau_i(k,p)$. We have also defined
\begin{eqnarray}
\nabla(k,p) &=& k^2 p^2 -(k \cdot p)^2 \,, \\
u(k,p) &=& 2 \, k\cdot p - \frac{\nabla(k,p)}{q^2} \,.
\end{eqnarray}
We are now in a position to go ahead and solve these coupled equations for any given vertex {\em ansatz}. We take up this task in perturbative and non-perturbative realms in the next section.

\section{Solving the gap equation}
\label{SECTION Solution gap equation}

In order to have a better intuitive grasp over the solutions of the gap equation, we start from perturbation theory. It is only natural to demand that any physically acceptable non-perturbative solution must reduce to its perturbative counterpart in the weak coupling regime.

\subsection{Perturbative solution}

Note that in the weak coupling regime, the gap equation is expected to reproduce the perturbative result for the electron propagator. In the chiral limit ($m_0=0$), the leading logarithmic expansion for the wave function renormalization in quenched RQED at one-loop order reads:
\begin{eqnarray}
F(k^2) = 1 + \frac{\alpha \varrho}{\pi} \left( \kappa - \phi \right) \log \left( \frac{k^2}{\Lambda^2} \right) + \frac{2}{3} \frac{\alpha \varrho}{\pi} \phi \,,
\label{Lead Log Wave Function}
\end{eqnarray}
where the last term in the above Eq.~(\ref{Lead Log Wave Function}), parameterized by the constant $\phi = \chi - 1/3$, spoils the power law behaviour of $F(k^2)$, Eq.~(\ref{F Power Law}). This term is spurious. It does not appear if we work in the dimensional regularization scheme. Just like in QED$_4$, its presence owes itself to the use of the cut-off regulator which we cannot avoid for the non-perturbative treatment presented in the next section. The good news is that just like in QED$_4$, we can cure the inadequacy of the cut-off regulator by an appropriate use of the WFGTI which ensures this spurious term is formally removed. It is achieved by redefining $\chi$ and $\kappa$ as prescribed through Eqs.~(\ref{Shift Chi},\ref{Shift Kappa}). To clarify how to achieve this, we focus on the survey of two special cases: \\ \\
{\bf Case 1:} Let us choose not to apply the WFGTI in any term of the kernel of the gap equation. It is equivalent to taking $\eta = \kappa$ in Eqs.~(\ref{Shift Chi},\ref{Shift Kappa}). It implies making the following shift:
\begin{eqnarray}
\chi & \rightarrow & \chi' = (1-\epsilon_e)\left( 1 - \xi \right) \,, \\
\kappa & \rightarrow & \kappa' = 0 \,,
\end{eqnarray}
yielding $\phi = -(\xi-1/3)/2$. \\ \\
{\bf Case 2:} Like the case of QED$_4$, we can be selective in applying the WFGTI on a redefined longitudinal part of the photon propagator such that $\phi = 0$. According to Eqs.~(\ref{Shift Chi},\ref{Shift Kappa}), it entails making the shift
\begin{eqnarray}
\chi & \rightarrow & \tilde{\chi} = 1/3 \,,  \label{Shif Chi Tilde}\\
\kappa & \rightarrow & \tilde{\kappa} = \epsilon_{e} - 2/3 + \xi (1 - \epsilon_{e}) \,. \label{Shif Kappa Tilde}
\end{eqnarray}
In this second case, the wave function renormalization exhibits the power law behaviour of Eq.~(\ref{F Power Law}) in the leading logarithmic approximation with  $\beta = \alpha (\xi -1/3) / (4 \pi)$ as stated before.

It is also worth noticing that for both the cases considered, we have
\begin{eqnarray}
\frac{1}{F(k^2)} - \frac{1}{F(p^2)} = - \frac{\alpha}{4 \pi} \left( \xi - \frac{1}{3} \right) \log \left( \frac{k^2}{p^2} \right) \,,
\label{Lead Log Result}
\end{eqnarray}
which proves to be a useful guide to express the one-loop corrections to the electron-photon vertex in terms of the electron dressing functions, as we study in section \ref{SECTION One-loop vertex}, thus helping us construct a non-perturbative {\em ansatz} for the electron-photon vertex.

So far, we have shown that ensuring the MR for the massless electron propagator, i.e., the power law behaviour of Eq.~(\ref{F Power Law}), at the one-loop level only requires a specific shifting of $\chi$ and $\kappa$, \textit{cf.} Eqs.~(\ref{Shif Chi Tilde},\ref{Shif Kappa Tilde}). However, two-loop and higher order perturbative expansions of the electron propagator and its gauge covariance will be sensitive to the choice of the vertex {\em ansatz}. For the sake of comparison, we compute the two-loop contributions to the electron dressing functions for the bare vertex ($\Gamma_{\mu} = \gamma_{\mu}$) and the BC-vertex, Eqs.~(\ref{Longitudinal vertex decomposition},\ref{longitudinal coefficients definitions}). For this purpose, we first compute the one-loop corrections to $F$ and ${\cal{M}}$ from Eqs.~(\ref{ProyM},\ref{ProyF}) for the case of a small mass (i.e., we neglect squared mass terms). The result is used recursively in the gap equation in order to compute two-loop contributions. Up to leading logarithmic terms, we arrive at the following expressions:
\begin{eqnarray}
F(k^2) &=& 1 + \beta \log \hspace{-.6mm} \left( \hspace{-.6mm} \frac{k^2}{\Lambda^2} \hspace{-.6mm} \right) + \frac{1}{2} \Phi_F \beta^2 \log^2 \hspace{-.6mm} \left( \hspace{-.6mm} \frac{k^2}{\Lambda^2} \hspace{-.6mm} \right), \label{2-loop F}\\
\frac{{\cal{M}}(k^2)}{m_0} &=&  1 + \lambda \log \hspace{-.6mm} \left( \hspace{-.6mm} \frac{\Lambda^2}{k^2} \hspace{-.6mm} \right) + \frac{1}{2} \Phi_{\cal{M}} \lambda^2 \log^2 \hspace{-.6mm} \left( \hspace{-.6mm} \frac{\Lambda^2}{k^2} \hspace{-.6mm} \right), \label{2-loop M}
\end{eqnarray}
where $\beta = \alpha \varrho \left( \kappa - \chi + 1/3 \right)/\pi$ in consonance with the one-loop expansion for $F$, Eq.~(\ref{Lead Log Wave Function}), and $\lambda = 8 \alpha \varrho / (3\pi)$. Moreover, for the bare vertex we have
\begin{eqnarray}
\Phi_F &=& \left( \kappa + \chi - 1/3 \right) / \left( \kappa - \chi + 1/3 \right) \,, \nonumber \\
\Phi_{\cal{M}} &=& \left\{ 23 + 6 \left( \kappa + \chi \right) - 3 \left( \kappa - \chi \right)  \left[ 8 + 3 \left( \kappa - \chi \right) \right] \right\}/32 \nonumber\,, \\
\label{Phi F&M Bare}
\end{eqnarray}
and for the BC-vertex
\begin{eqnarray}
\Phi_F &=& \left( \kappa - \chi + 4/3 \right) / \left( \kappa - \chi + 1/3 \right) \,, \nonumber \\
\Phi_{\cal{M}} &=& \left\{ 40 - 3 \left( \kappa - \chi \right)  \left[ 1 + 3 \left( \kappa - \chi \right) \right] \right\}/32 \,.
\label{Phi F&M BC}
\end{eqnarray}
It is important to notice from Eqs.~(\ref{2-loop F},\ref{2-loop M}) that a power law solution for $F$ and ${\cal{M}}$, \textit{cf.} Eq.~(\ref{Lead Log Wave Function}), demands $\Phi_F = \Phi_{\cal{M}} = 1$. However, there is no shift in $\chi$ and $\kappa$ that ensures this requirement. Moreover, for both bare and BC vertices, the mass function picks up a gauge dependence at second order in perturbation theory via $\kappa \sim \xi$ in $\Phi_{\cal{M}}$. These results show that even the full longitudinal vertex fails to ensure the local gauge covariance of the electron propagator and highlights the necessity of incorporating a transverse component for the electron-photon vertex.

In order to elucidate the analytical structure of the transverse vertex and its impact on the gauge covariance of the electron propagator, it is worth realizing that the leading logarithmic contributions to $F$ and ${\cal{M}}$ at two-loops, Eqs.~(\ref{2-loop F},\ref{2-loop M}), arise from integrals of the form
\begin{eqnarray}
\int_{k}^{\Lambda} \frac{dp}{p} &=& \frac{1}{2} \log \hspace{-.6mm} \left( \hspace{-.6mm} \frac{\Lambda^2}{k^2} \hspace{-.6mm} \right) \,, \nonumber \\
\int_{k}^{\Lambda} \frac{dp}{p} \log \hspace{-.6mm} \left( \hspace{-.6mm} \frac{p^2}{\Lambda^2} \hspace{-.6mm} \right) &=& \frac{1}{4} \log^2 \hspace{-.6mm} \left( \hspace{-.6mm} \frac{\Lambda^2}{k^2} \hspace{-.6mm} \right) \,, \nonumber
\end{eqnarray}
which in turn correspond to evaluating the integrals in the so-called asymptotic regime where $p^2 \gg k^2 \gg m^2_0$. Since the two-loop contributions to the electron propagator are determined by the one-loop corrections of the electron-photon vertex, in particular its transverse pieces, we reaffirm that not only a proper treatment of the photon propagator and an adequate use of the WFGTI are essential to study the gap equation but an inclusion of a refined transverse vertex is also required to incorporate the constraints of gauge covariance and perturbation theory as advocated before as well.



\begin{figure}[!ht]
    \centering
    \includegraphics[scale=.26]{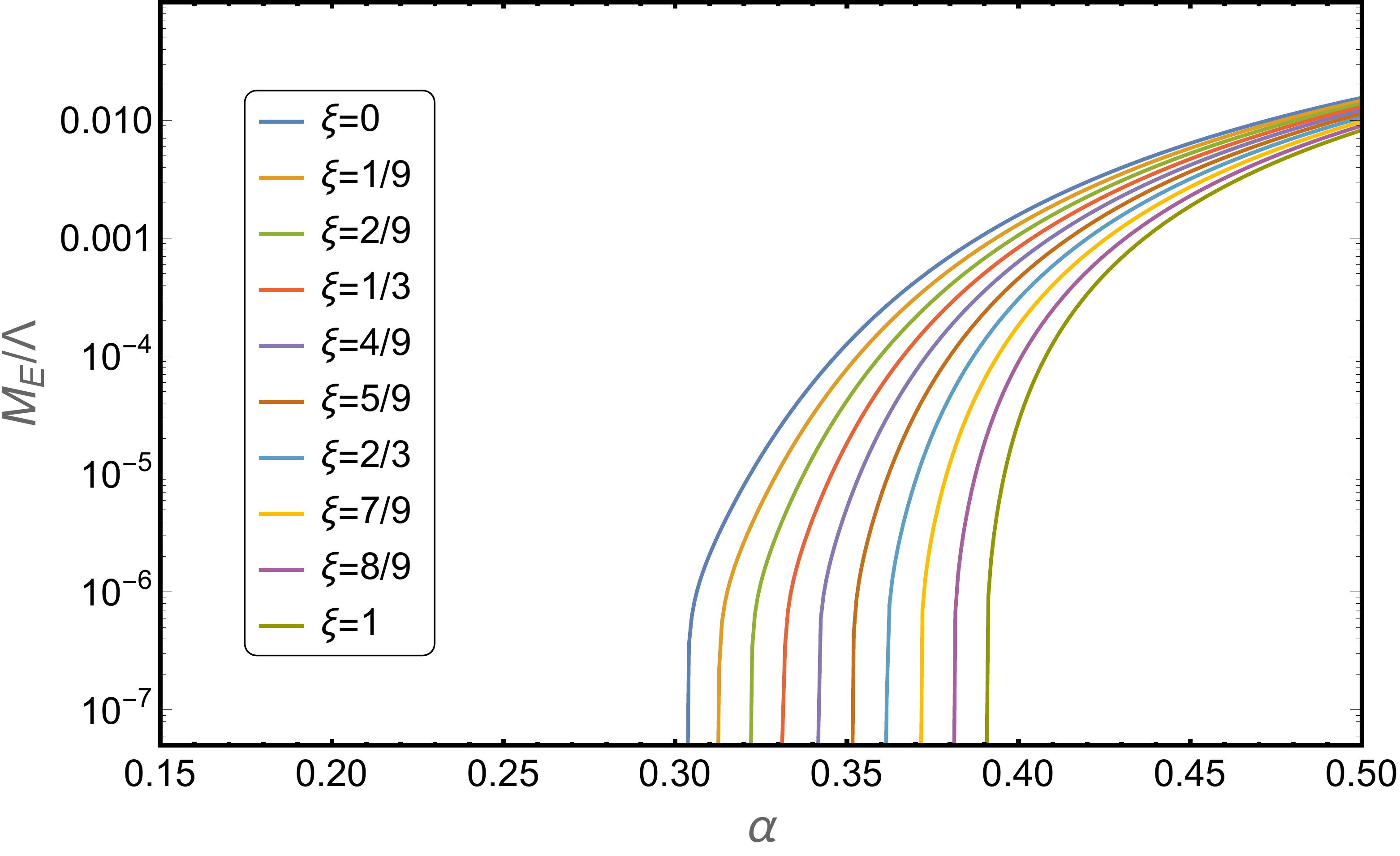}
    \caption{Dynamically generated dimensionless Euclidean mass $M_E/\Lambda$ as a function of the electromagnetic coupling $\alpha$ for the bare vertex {\em ansatz} for different values of the covariant gauge parameter $\xi$. The gauge dependence of $M_E/\Lambda$ and $\alpha_c$ is noticeably sizeable even in a small interval of $\xi=$, i.e., (0,1).}
\label{BV-gauge-dependence}
\end{figure}

\subsection{Non-perturbative solution and the DCSB}
\label{SECTION DCSB}

In order to compute non-perturbative solutions of the electron SDE or the gap equation, Eq.~(\ref{gap equation}), it is necessary to choose an {\em ansatz} for the electron-photon vertex defined through Eqs.~(\ref{Ball-Chiu vertex decomposition},\ref{Longitudinal vertex decomposition},\ref{transverse vertex structure},\ref{transverse basis definition}). We
compute dynamically generated massive solutions for the electron propagator in the chiral limit, i.e., when the bare electron mass $m_0=0$. In this case, the electron mass generated is entirely dynamical in nature. In particular, we compute $M_E={\cal M}(k^2)$ which can be interpreted as the Euclidean {\em mass}. Note that the only mass scale available in quenched RQED is the ultraviolet regulator $\Lambda$. The dynamically generated mass $M_E$ is naturally proportional to it. Therefore, we choose to plot $M_E/\Lambda$ as a function of the electromagnetic coupling $\alpha$. At $\alpha=\alpha_c$, a DCSB solution emerges, bifurcating away from the perturbative solution $M_E=0$ which corresponds to the fact that no electron mass is generated at any order in perturbation theory if we start from the bare mass $m_0=0$.
As $\alpha_c$ separates the phase of massless electrons from that of massive electrons, we expect it to be gauge invariant just as the Euclidean mass itself\footnote{It is the Minkowski pole mass which should strictly be gauge invariant. However, as $M_E$ lies close to $M(0)$, we would expect it be at least nearly gauge invariant.}. Note that we only depict results after the appropriate use of the WFGTI in the kernel of the gap equation. To start with, we compare the results of the following two vertices:

\begin{figure}[!ht]
    \centering
    \includegraphics[scale=.26]{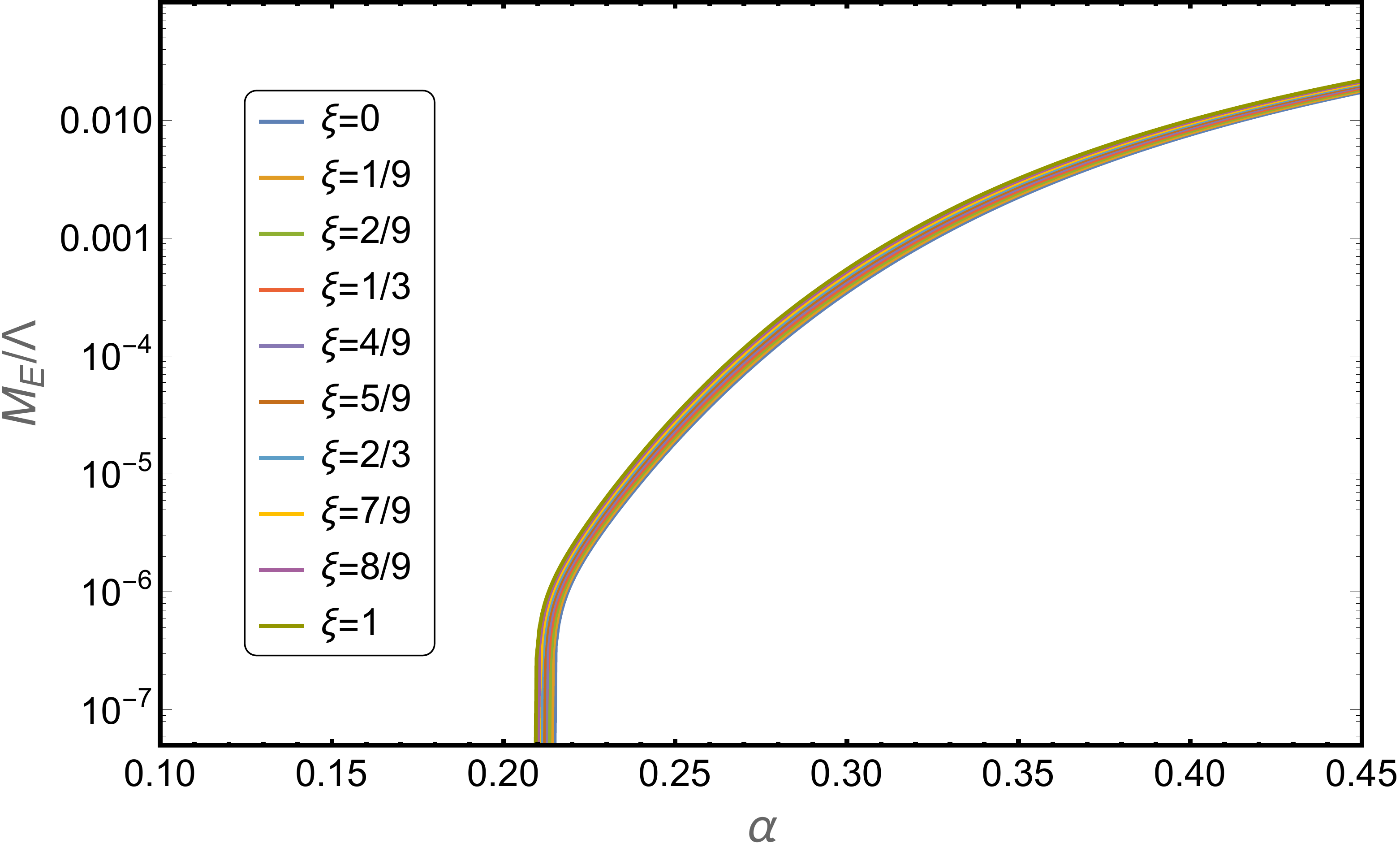}
    \caption{Dynamically generated dimensionless Euclidean electron mass $M_E/\Lambda$ as a function of the electromagnetic coupling $\alpha$ for our proposed vertex construction for different values of the covariant gauge parameter $\xi$. Both $M_E/\Lambda$ and $\alpha_c$ are now practically gauge independent!}
\label{FV-gauge-independence}
\end{figure}

\begin{itemize}
    \item \textbf{Bare}, with $\lambda_1=1$ and $\lambda_{2,3} = \tau_{1,2,...,8} = 0$.
     \item \textbf{Our {\em ansatz}}, with $\lambda_{1,2,3}$ defined in Eq.~(\ref{longitudinal coefficients definitions}) and all $\tau$'s set to zero except $\tau_6$ defined in Eq.~(\ref{Tau 6 Asymptotic}) with $a_6=-1/2$.
     \label{FV-gauge-dependence}
\end{itemize}
Fig.~(\ref{BV-gauge-dependence}) clearly depicts the fact that the bare vertex results in considerable amount of undesirable gauge dependence in the Euclidean mass $M_E$ and $\alpha_c$. We now repeat the exercise for our proposed vertex {\em ansatz} which not only agrees with the asymptotic one-loop result for the full vertex but also guarantees the wave-function renormalization to be MR in the leading logarithmic approximation, see Fig.~(\ref{FV-gauge-independence}).

\begin{figure}[!ht]
    \centering
    \includegraphics[scale=.26]{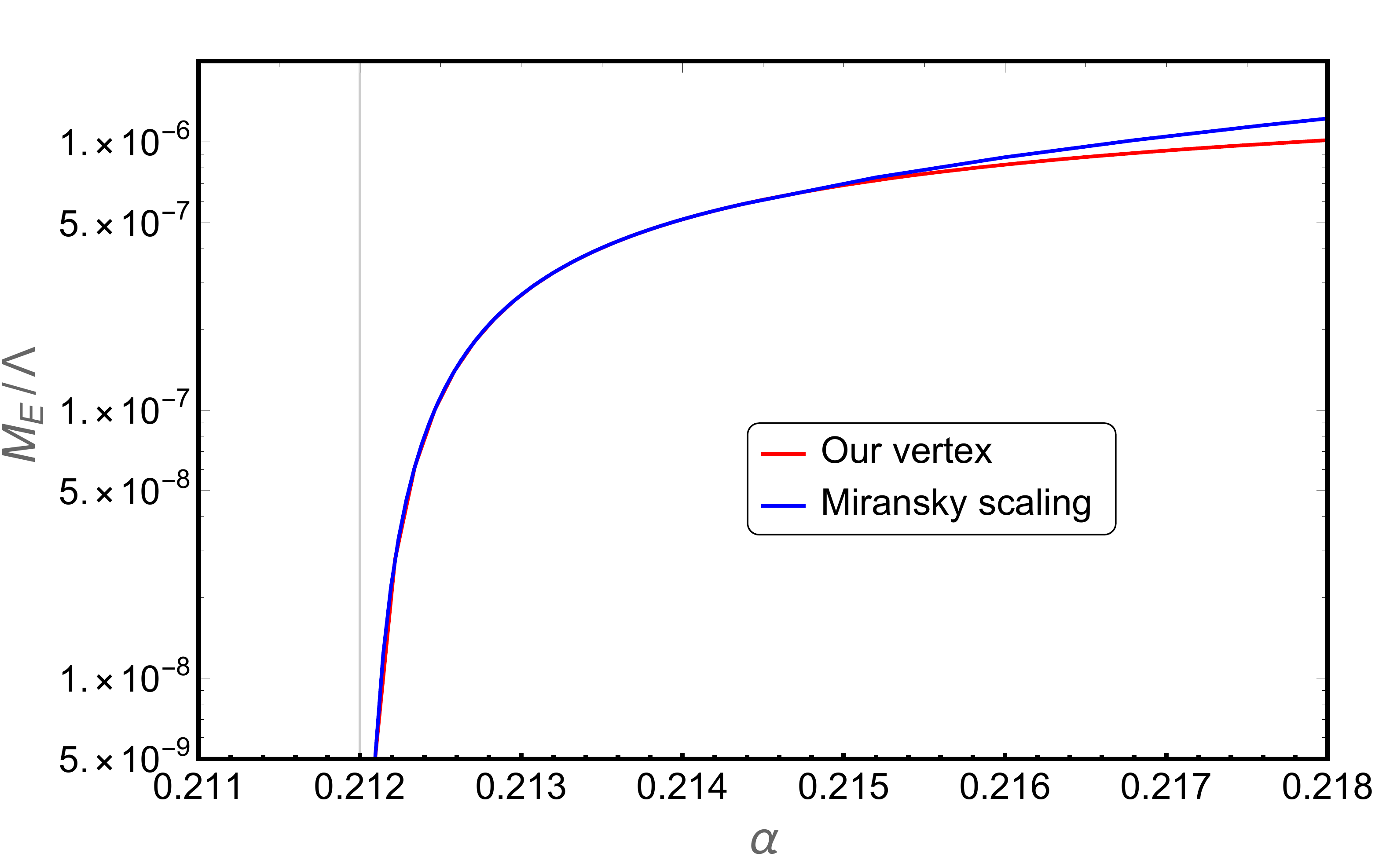}
    \caption{Miransky Scaling using our vertex for $\xi = 1/3$.} \label{Miransky scaling}
\end{figure}

The gauge independence is almost magically instated! We  find that our choice of the vertex yields $\alpha_c (\xi) \approx 0.212$.
We confirm that the $\alpha$-dependence of $M_E/\Lambda$ in the close vicinity of $\alpha_c$ satisfies Miransky scaling law, see Fig.~(\ref{Miransky scaling}):
 \begin{eqnarray}
\frac{M_E}{\Lambda} = {\rm Exp} \left[ -\frac{ \pi \kappa_1}{\sqrt{\alpha/\alpha_c -1}} + \kappa_2\right] \,,
 \end{eqnarray}
where $\kappa_1= 0.051$ and $\kappa_2= -12.857$.

\begin{figure}[!ht]
    \centering
    \includegraphics[scale=.26]{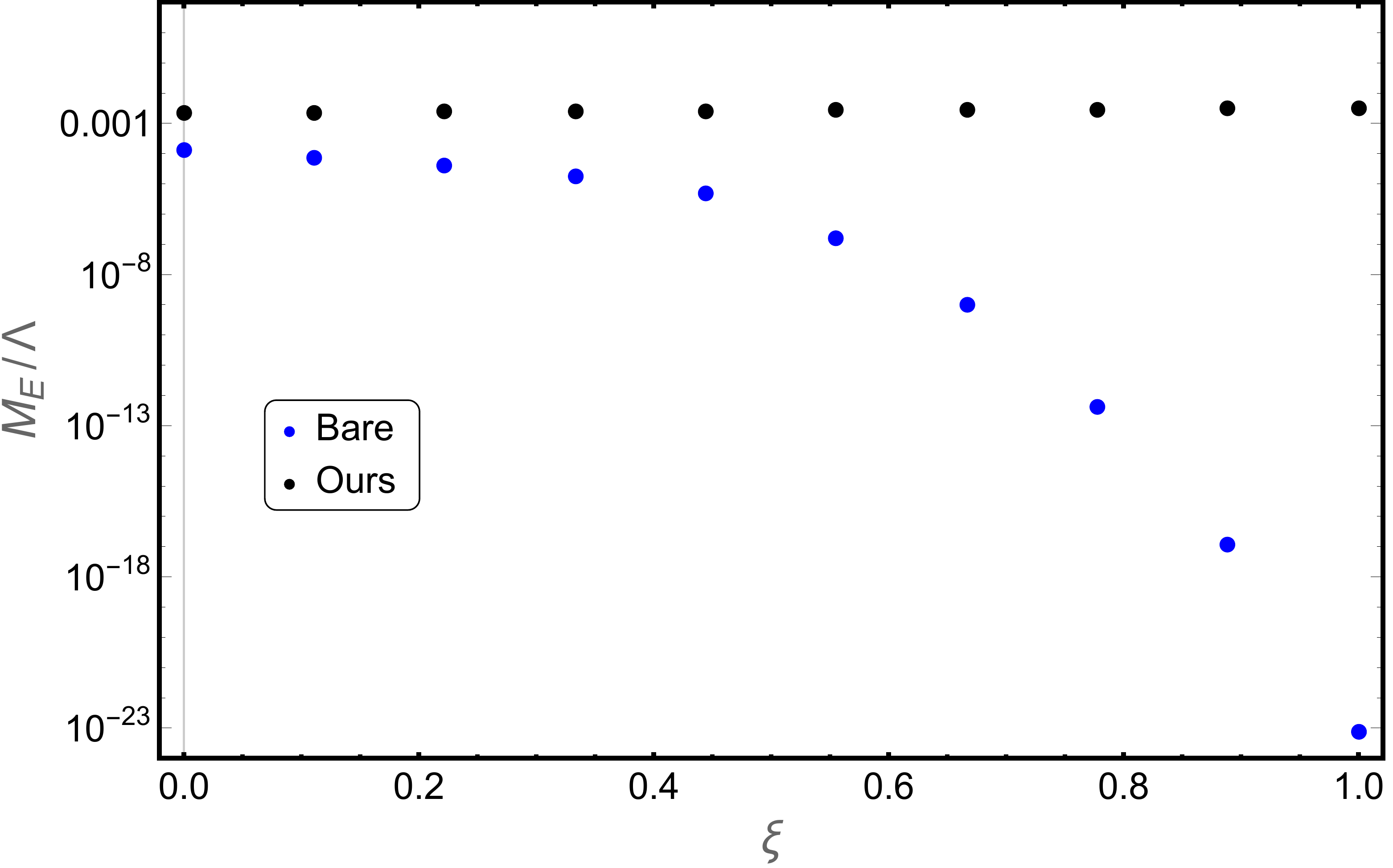}
    \caption{Dynamically generated dimensional Euclidean mass as  a function of the covariant gauge parameter $\xi$ for the bare vertex and our proposed vertex {\em ansatz}. We choose $\alpha=0.35$ to draw the plot.} \label{M vs Xi}
    \label{Fig Mass Gauge}
\end{figure}

\begin{figure}[!ht]
    \centering
    \includegraphics[scale=.26]{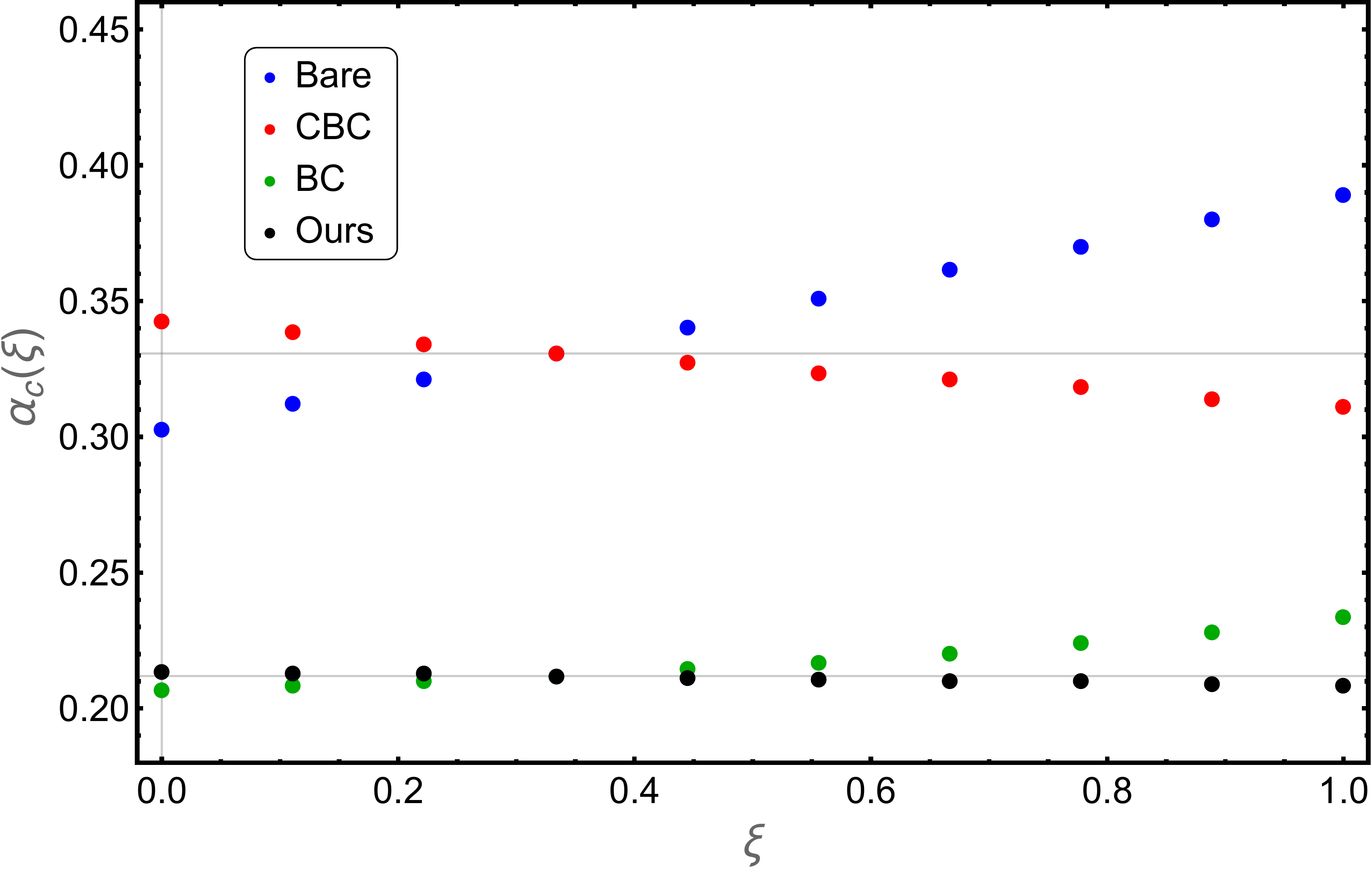}
    \caption{The critical coupling $\alpha_c$ as a a function of the covariant gauge parameter $\xi$ for four different choices of the vertex {\em Ans$\ddot{a}$tze}, mentioned in the text. } \label{Fig-gauge-dependence}
\end{figure}
At this point, it appears worth its while to plot the dynamical mass $M_E/\Lambda$ as a function of the covariant gauge parameter $\xi$, Fig.~({\ref{Fig Mass Gauge}}). We choose $\alpha=0.35$ merely as a representative value to bring out the stark difference between the two truncation schemes as regards the gauge dependence of this physical observable. It is remarkable to note the potency of our proposal in this connection.

We also plot $\alpha_c(\xi)$ to have a more quantitative insight into its gauge (in)dependence, Fig.~(\ref{Fig-gauge-dependence}). Moreover, one might consider it illustrative to also compare the results with the following two vertices occasionally adopted in literature:
\begin{itemize}
    \item \textbf{Central Ball-Chiu} (CBC), with $\lambda_1$ defined in Eq.~(\ref{longitudinal coefficients definitions}) and $\lambda_{2,3} = \tau_{1,2,...,8} = 0$.
    \item \textbf{Ball-Chiu} (BC), with $\lambda_{1,2,3}$ defined in Eq.~(\ref{longitudinal coefficients definitions}) and $\tau_{1,2,...,8} = 0$.
\end{itemize}



Notably the bare vertex and the CBC-vertex perform almost equally badly! The BC-vertex reduces the gauge dependence. However, our proposal renders $\alpha_c$ practically gauge independent.
These observations naturally lead us to conclude the article on a positive note.

\section{Conclusions and Perspectives}

It is quite satisfactory to observe that a truncation of the infinite tower of SDEs which respects key features of the underlying quantum field theory of RQED, namely its discrete symmetries, matching with perturbation theory in the domain of weak coupling and the MR of the massless electron propagator expectedly leads to practically gauge invariant results, within the interval $(0,1)$ of the gauge parameter $\xi$ for the observales under study, i.e., the dynamically generated Euclidean mass and the critical coupling which marks the onslaught of DCSB.
What is remarkable is that only one transverse form factor $\tau_6(k,p)$ suffices to construct our vertex {\em ansatz} and to achieve the goal we set ourselves.
Having constructed a reliable truncation of the infinite tower of the SDEs, we believe we are now in a pole position to study the physical properties of Graphene and other relevant materials of interest within this formalism. This is for future.

\label{SECTION Conclusions and Perspectives}

 \section*{Acknowledgments}
This work is partly funded by CONACyT through the post doctoral scholarship for L.~Albino. A.~Bashir acknowledges the CIC(UMSNH) grant 4.10.
A.~J.~Mizher is grateful to the financial support from FAPESP under fellowship number 2016/12705-7 and  A.~Raya to the funding from the CONACYT Project FORDECYT-PRONACES/61533/2020.

\newpage\bibliography{References}

\end{document}